\documentclass[acmtog]{acmart}

\copyrightyear{2026}
\acmYear{2026}
\setcopyright{cc}
\setcctype{by}

\acmConference[SA Conference Papers '26]
{SIGGRAPH Asia 2026 Conference Papers}
{December 01--04, 2026}
{Kuala Lumpur, Malaysia}

\acmBooktitle{SIGGRAPH Asia 2026 Conference Papers (SA Conference Papers '26), December 01--04, 2026, Kuala Lumpur, Malaysia}

\acmDOI{10.1145/3829340.3842262}
\acmISBN{979-8-4007-2842-6/2026/12}

\newcommand{\YH}[1]{{\color{violet}{{[Oh: #1]}}}}

\newcommand{\TODO}[1]{{{\textbf{\color{red}{[TODO: #1]}}}}}

\setcitestyle{square}

\usepackage{amsbsy}
\usepackage{amsmath}

\usepackage{nicefrac}

\usepackage[ruled]{algorithm2e}

\SetAlFnt{\small}
\SetAlCapFnt{\small}
\SetAlCapNameFnt{\small}
\SetAlCapHSkip{0pt}
\IncMargin{-\parindent}

\usepackage{dblfloatfix}

\usepackage{array}
\usepackage{makecell}
\usepackage{subcaption}
\usepackage{enumitem}

\begin{document}

\title{Real-world Polarimetric Environment Map Dataset}

\begin{teaserfigure}
    \includegraphics[width=\linewidth]{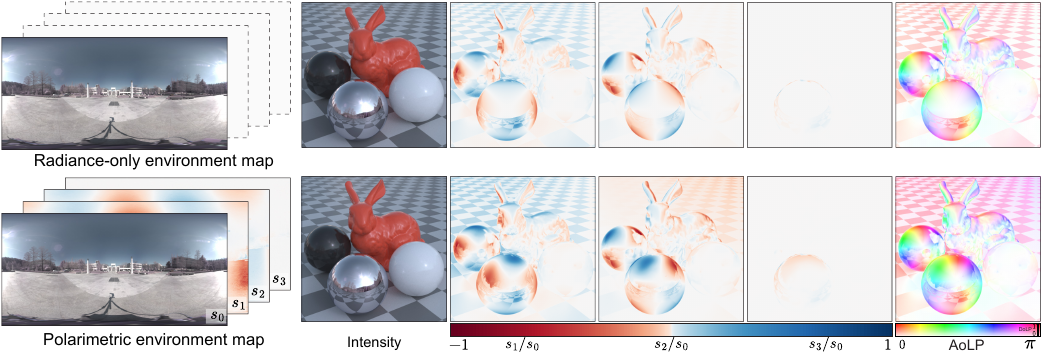}
    \caption{\textbf{Radiance-only environment map vs. polarimetric environment map.} Top: Conventional environment maps encode only radiance and implicitly assume unpolarized light, preventing them from reproducing the polarization effects present in real-world lighting. Bottom: Our polarimetric environment maps provide full-sphere Stokes vectors, enabling polarized image-based lighting and producing realistic renderings of natural polarized illumination.}
    \label{fig:teaser}
\end{teaserfigure}

\author{Yonghee Oh}
\authornote{Both authors contributed equally to this research.}
\email{dhdydgml1019@postech.ac.kr}
\affiliation{%
  \institution{POSTECH}
  \country{South Korea}
}

\author{Ryota Maeda}
\authornotemark[1]
\email{maeda.ryota.elerac@gmail.com}
\affiliation{%
  \institution{POSTECH}
  \country{South Korea}
}

\author{Yunseong Moon}
\email{yunseong0518@postech.ac.kr}
\affiliation{%
  \institution{POSTECH}
  \country{South Korea}
}

\author{Seung-Hwan Baek}
\email{shwbaek@postech.ac.kr}
\affiliation{%
  \institution{POSTECH}
  \country{South Korea}
}

\authorsaddresses{}

\begin{abstract}

Environment maps are a standard representation of illumination, enabling realistic image-based lighting in modern rendering systems. However, existing environment maps typically encode only RGB radiance, omitting polarization, an intrinsic property of light that is imperceptible to the human eye but informative about surface shape and material properties. This omission precludes the physically accurate simulation of polarized light transport under in-the-wild illumination conditions, heavily limiting the development and evaluation of robust polarization-based scene reconstruction and understanding methods.
We address this gap by introducing a real-world polarimetric environment map dataset, each providing a full-sphere panorama with per-pixel high-dynamic-range RGB Stokes vectors. To acquire the dataset, we design and calibrate a dedicated imaging system and develop a polarization-aware panorama-stitching pipeline. In particular, we introduce a frame-field representation for perspective cameras, together with singularity-aware blending, to enable smooth polarimetric panorama stitching without singularity artifacts.
Using this system, we capture 100 polarimetric environment maps across diverse real-world scenes. We further develop a custom rendering plugin and demonstrate close agreement between real and rendered observations. Finally, we analyze the polarization state of the sky across different times and weather conditions, and demonstrate its practical impact on a downstream shape-from-polarization task.
Our dataset and code are available on our project page\footnote{\url{https://yongheeoh.github.io/PolarimetricEnvironmentMap/}}.

\end{abstract}



\maketitle
\newcommand{\vect}[1]{\mathbf{#1}}
\newcommand{\mat}[1]{\mathbf{#1}}

\newcommand\note[1]{\textcolor{red}{#1}}

\newcommand{\psf}{\rho}

\newcommand{\argmin}[1]{\stackrel[\{ #1 \}]{}{\textrm{arg min}}}
\newcommand{\minimize}[1]{\stackrel[\{ #1 \}]{}{\textrm{minimize}}}

\section{Introduction}
\label{sec:intro} 
High-dynamic-range (HDR) environment maps have become a de facto representation of real-world illumination in computer graphics and vision. Since Debevec’s seminal introduction of image-based lighting~\cite{debevec1998rendering}, environment maps have been widely used to reproduce realistic illumination in physically based rendering systems. Modern HDR environment map collections—including the Laval HDR database~\cite{gardner2017learning, hold2019deep, bolduc2023beyond}, the UrbanSky database~\cite{klotz2025minimal}, and large-scale public repositories such as PolyHaven~\cite{PolyHaven}—now provide thousands of high-resolution panoramas spanning diverse lighting conditions. These datasets, however, encode only intensity and omit the polarization.

Polarization, a wave property of light, has become a powerful cue for geometry and material analysis~\cite{wolff1990polarization, wolff1991constraining, miyazaki2003polarization, cui2017polarimetric, lei2020polarized, kalra2020deep, dave2022pandora, baek2021polarimetric}. Consequently, physically accurate polarimetric rendering is increasingly important not only for evaluating polarization-based methods but also for generating synthetic supervision and analysis data for future polarimetric vision algorithms. 
A standard rendering framework for polarization uses the Stokes–Mueller formalism~\cite{collett2005field}, and modern physically based renderers such as Mitsuba~3~\cite{Mitsuba3} now support polarization-aware rendering. The demand for realistic polarimetric rendering has driven the acquisition of polarimetric bidirectional reflectance distribution functions (pBRDFs) of real-world materials~\cite{baek2020image, moon2025hyperspectral}, enabling data-driven characterization of polarization-dependent reflectance.

In contrast, the environmental illumination side of the polarimetric rendering remains underexplored. Nearly all existing work assumes completely unpolarized environmental illumination~\cite{dave2022pandora, lyu2024sfpuel} or relies on synthetically generated polarized environment maps~\cite{yi2024spin}. Such simplifications limit the faithful simulation of polarized light transport under natural polarized illumination.

In this paper, we introduce a real-world polarimetric environment map dataset. To our knowledge, this is the first released dataset of real-world, full-sphere, HDR RGB-Stokes environment maps designed for polarization-aware image-based lighting. The dataset covers diverse indoor and outdoor environments across various times, enabling physically grounded polarimetric rendering that was previously not possible with radiance-only environment maps.

To acquire polarimetric environment maps, we construct an imaging system consisting of a full-Stokes camera mounted on a motorized rotation stage. We develop a dedicated polarization-aware panorama-stitching pipeline. Our method defines a perspective measurement frame suitable for wide-field-of-view (FoV) polarimetric images and introduces a singularity-aware blending strategy based on a continuous frame-free representation.

We also develop a custom plugin for Mitsuba~3~\shortcite{Mitsuba3} that supports rendering with our polarimetric environment maps. The rendered polarization images closely match real captured observations, validating both our acquisition system and rendering pipeline.
Using time-varying polarimetric environment-map sequences captured from a fixed viewpoint, we analyze polarization patterns in the sky. 
We further demonstrate the utility of the dataset through a controlled SfP experiment isolating polarization from radiance.

In summary, our contributions are:
\begin{itemize}
    \item We present the first dataset of real-world HDR polarimetric environment maps, comprising 100 full-sphere panoramas captured, together with rendering code.
    \item We introduce a polarization-aware panorama-stitching method that addresses polarization-specific challenges, including measurement frame definition for wide-FoV cameras and singularity artifacts.
    \item We analyze real-world sky polarization by comparing our measurements with analytical sky models and use a controlled experiment to demonstrate the effect of real-world polarized illumination.
\end{itemize}

Our dataset opens a new avenue for physically realistic polarimetric illumination and provides a foundation for developing and evaluating polarization-based visual computing.

\section{Related Work}
\label{sec:related}
\paragraph{Environment Maps.}
Classical environment map acquisition dates back to Debevec’s seminal work~\cite{debevec1998rendering}, which acquires real-world illumination by photographing a mirrored sphere under multi-exposure brackets.
Subsequent datasets scaled this paradigm to large collections of real-world RGB HDR panoramas, including the Laval Indoor and Outdoor HDR datasets~\cite{gardner2017learning, hold2019deep, bolduc2023beyond}, captured from multiple viewpoints, and the UrbanSky dataset~\cite{klotz2025minimal}, captured using a $360^\circ$ camera. These datasets provide large-scale measurements of real-world illumination and have served as important ground truth for realistic rendering, illumination analysis, and inverse-rendering benchmarks.
Beyond RGB imaging, several works have explored the acquisition of multi-spectral~\cite{tominaga2010spectral, hirai2018high} and hyperspectral~\cite{morimoto2019hyperspectral} environment maps for spectral rendering and analysis. 
Despite extensive prior work, existing datasets are limited to RGB or spectral radiance alone.


\paragraph{Polarimetric Rendering.}
Polarimetric rendering extends traditional light-transport simulation by employing the Stokes–Mueller formalism to represent both the polarization state of illumination and the polarization-dependent reflectance of materials.
The polarized path tracer was introduced by \citet{wilkie2012polarised}, and subsequent work established bidirectional extensions~\cite{jarabo2018bidirectional, mojzik2016bi}.
Polarization-dependent material interactions are modeled using pBRDFs through parametric formulations~\cite{baek2018simultaneous, ichikawa2023fresnel} or direct measurements of real-world materials~\cite{baek2020image, moon2025hyperspectral}.
More recent advances utilize spherical harmonics to enable efficient frequency-domain simulation~\cite{yi2024spin}.
Modern rendering systems, such as Mitsuba~3~\cite{Mitsuba3}, provide full-Stokes light transport rendering, yet the community still continues to rely on unpolarized or analytically synthesized illumination due to the absence of a real-world polarized environment map dataset. 

\paragraph{Applications of Polarimetric Imaging.}

Polarization provides complementary cues for diverse inverse imaging problems. For shape recovery, it has been combined with depth sensing~\cite{kadambi2015polarized}, multiview geometry~\cite{cui2017polarimetric}, and normal stereo~\cite{fukao2021polarimetric}; Shape from Sky~\cite{ichikawa2021shape} further exploits polarized skylight. Beyond shape recovery, polarization has enabled HDR imaging~\cite{zhou2023polarization, xie2023snapshot}, road detection and glass segmentation~\cite{li2020full, mei2022glass}, low-light enhancement, demosaicing, super-resolution, and deblurring~\cite{zhouaaai2023polarization, zhou2025pidsr, zhou2025learning}, as well as non-line-of-sight imaging~\cite{tanaka2020polarized}. Recent methods also incorporate polarization into Gaussian-based scene reconstruction~\cite{shan2026polarguide, wu2026phygap} and RGB-to-polarization estimation~\cite{lin2026rgb}. These applications motivate real-world full-sphere polarized illumination for physically grounded rendering and controlled evaluation.

\paragraph{Polarization Datasets.}
With the growing importance of data-driven polarimetric imaging and the commercial availability of linear polarization sensors~\cite{yamazaki2016four}, numerous polarization datasets have been introduced for polarization-based vision tasks.
Task-specific datasets have been developed for shape-from-polarization \cite{ba2020deep}, inverse-rendering~\cite{dave2022pandora, han2024nersp, li2025neisf++}, reflection removal~\cite{yao2025polarfree}, material segmentation~\cite{liang2022multimodal}, demosaicing~\cite{qiu2021linear, rahman2025polarization} and burst imaging~\cite{hwang2025benchmarking}. 
Beyond linear measurements, \citet{jeon2024spectral} present a full-Stokes polarimetric image dataset containing both linear and circular polarization measurements.
However, existing datasets do not include polarimetric illumination. Our work addresses this gap by introducing the first polarimetric environment map dataset.

\section{Background on Polarization}
\label{sec:background}

\paragraph{Stokes-Mueller Formalism}
The Stokes-Mueller formalism is a mathematical framework describing the polarization state of light and its transformation in interaction with materials~\cite{baek2023polarization}. 
The polarization state of light is represented by a Stokes vector $\mathbf{s} = [s_0, s_1, s_2, s_3]^\intercal \in \mathbb{R}^{4 \times 1}$, where $s_0$ corresponds to the total intensity of the light. The parameters $s_1$ and $s_2$ quantify the intensity differences of linearly polarized light at $0^\circ/90^\circ$ and $45^\circ/-45^\circ$ respectively, while $s_3$ represents the intensity difference between right- and left-circularly polarized light. 
The transformation of the Stokes vector by interacting with matter is described by the Mueller matrix $\mathbf{M} \in \mathbb{R}^{4\times4}$. The relationship between the input $\mathbf{s}_\text{in}$ and output Stokes vectors $\mathbf{s}_\text{out}$ is then expressed as $\mathbf{s}_\text{out} = \mathbf{M}\mathbf{s}_\text{in}$.


\paragraph{Reference Frame of the Stokes-vector}
A Stokes vector is defined with respect to a local orthonormal frame $\mathbf{F} = [\mathbf{x}, \mathbf{y}, \mathbf{z}] \in \mathbb{R}^{3 \times 3}$, in which $\mathbf{z}$ is aligned with the direction of light propagation and $(\mathbf{x}, \mathbf{y})$ span the transverse plane. Multiple conventions exist for constructing such a frame: Mitsuba~3 adopts Duff's construction~\cite{duff2017building}, while Yi et al.\ introduce a $\theta\phi$-frame tailored to spherical parameterizations~\cite{yi2024spin}. Because the Stokes parameters are frame-dependent, the numerical values of $\mathbf{s}$ change under a rotation of $\mathbf{F}$. Let $\mathbf{F}'$ denote the frame obtained by rotating $\mathbf{F}$ about the propagation direction $\mathbf{z}$ by an angle $\vartheta$. The Stokes vector expressed in $\mathbf{F}'$ is then given by the rotation Mueller matrix
\begin{equation}
    \label{eq:rotation_mueller}
    \mathbf{s} \leftarrow
    \mathbf{C}_{\mathbf{F}\to\mathbf{F}'} \ \mathbf{s}, \quad \text{where }
    \mathbf{C}_{\mathbf{F} \to \mathbf{F}'} =  
    \begin{bmatrix}
        1 & 0 & 0 & 0 \\
        0 & \cos 2\vartheta & \sin 2\vartheta & 0 \\
        0 & -\sin 2\vartheta & \cos 2\vartheta & 0 \\
        0 & 0 & 0 & 1
    \end{bmatrix}.
\end{equation}
This matrix rotates the linear polarization components $s_1$ and $s_2$ by twice the frame rotation angle $2\vartheta$, while leaving $s_0$ and $s_3$ unchanged.

\paragraph{Singularity Problem and S2L2}
Such rotations are essential for aligning Stokes vectors defined in different local frames. However, any continuous frame field on the sphere must contain singularities, which can produce discontinuities and sign-flipping artifacts in full-sphere environment maps. 
To avoid these artifacts, \citet{yi2025frame} introduced S2L2, a frame-free Stokes representation continuous over all directions. Our polarimetric panorama stitching pipeline builds on S2L2 and is therefore free of frame-singularity artifacts.


\paragraph{Visualization of Polarization Images}
For visualization, we convert the Stokes vector into interpretable polarization descriptors, including the angle of linear polarization (AoLP), the degree of linear polarization (DoLP), the degree of polarization (DoP) $\rho \in [0, 1]$, and the ellipticity angle $\chi \in [-\pi/4, \pi/4]$. Specifically, AoLP and DoLP are computed as $\psi = 0.5\operatorname{atan2}(s_2,s_1)$, with $\psi$ wrapped to $[0,\pi)$, and $\rho_L = \sqrt{s_1^2+s_2^2}/s_0$, respectively. We further represent the ellipticity angle using the type of polarization (ToP) and the chirality of polarization (CoP). Throughout this paper, we follow the standardized color scheme of \citet{wilkie2010standardised} for visualizing Stokes-vector images.

\section{Acquisition of Spherical Stokes Images}
\label{sec:acquisition}

\paragraph{Representation}

Polarimetric environment maps extend conventional HDR environment maps, which encode radiance as a function of direction, by additionally recording the polarization state of incident illumination. 
Specifically, we define each pixel of a polarimetric environment map as a mapping from the physical propagation direction of light $\boldsymbol{\omega} \in \mathbb{S}^2$ to a Stokes vector $\mathbf{s}_\mathrm{env}(\boldsymbol{\omega})$ expressed in a local orthonormal frame $\mathbf{F}_\mathrm{env}(\boldsymbol{\omega})$.


\paragraph{Imaging Hardware}

\begin{figure}[t]
    \centering
    \includegraphics[width=\linewidth]{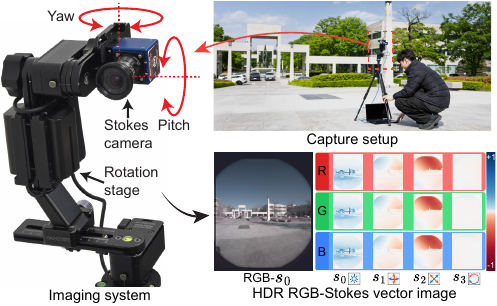}
    \caption{\textbf{Imaging system and capture setup for polarimetric environment map acquisition.} The system consists of a single-shot Stokes camera mounted on an automated dual-axis rotation stage that controls yaw and pitch. For each rotation angle, the camera captures exposure-bracketed images and merges them into an HDR RGB--Stokes image.}
    \label{fig:imaging_system}
\end{figure}

To acquire densely sampled polarimetric observations over the full sphere, we construct a dedicated capture system comprising (i) a single-shot Stokes camera and (ii) an automated dual-axis panoramic rotation stage. Figure~\ref{fig:imaging_system} shows our imaging system for polarimetric environment map acquisition. 
\begin{itemize}[leftmargin=*]
    \item The Stokes camera~\cite{tu2020division, jeon2024spectral} integrates custom micro retarders on top of a commercially available RGB polarization sensor (Sony, IMX250MYR). This optical architecture enables per-pixel measurement of both linear and circular polarization components in each RGB channel. 
    The sensor resolution is 2448$\times$2048, and each pixel stores an RGB Stokes vector. For the imaging lens, we use a wide perspective lens (Edmund Optics, UC Series 4\,mm focal length) with an approximately $100^\circ$ FoV.
    
    \item The rotation stage (Nodal Ninja F9921) provides motorized yaw and pitch control, enabling accurate and efficient sampling of viewing directions over the sphere. In addition, the stage offers manual translation adjustment along two axes to align the camera with the stage rotation center. Prior to capture, we manually adjust this translation so that the camera rotates about its optical center (no-parallax point), thereby minimizing view-dependent misalignment during panorama stitching.
\end{itemize}


\paragraph{Calibration}
We perform both one-time and per-scene calibration.
For one-time calibration, we first estimate the camera intrinsics matrix and distortion parameters using Zhang's method~\shortcite{zhang2002flexible}. We then calibrate the polarimetric response of our Stokes camera so that the measured polarization states match known incident states generated by a linear polarizer and a quarter-wave plate. We also calibrate spatially varying light attenuation, including lens vignetting and sensor dust, by capturing a uniformly illuminated white flat-field target. The resulting flat-field image is used to compensate for per-pixel intensity falloff in all captured images. Finally, we manually create a validity mask to exclude pixels outside the effective field of view.
Unlike these one-time calibrations, we perform per-scene calibration. Specifically, we estimate white-balance gains by capturing a color checker. We also estimate the camera rotation matrix for each view by solving a rotation-only pose optimization problem from feature correspondences. 
Details of the calibration procedure are provided in the supplementary material.



\paragraph{Per-view Capture and Pre-processing}
For each scene, we capture 18 views by uniformly sampling the viewing sphere using six yaw angles and three pitch angles.
We then apply a sequence of preprocessing steps, including HDR fusion, vignetting correction, demosaicing, Stokes-vector estimation, white balancing, and color correction. This process yields one HDR RGB--Stokes image for each view.

\section{Polarimetric Panorama Image Stitching}
\label{sec:stitching}

\begin{figure*}[t]
    \centering
    \includegraphics[width=\linewidth]{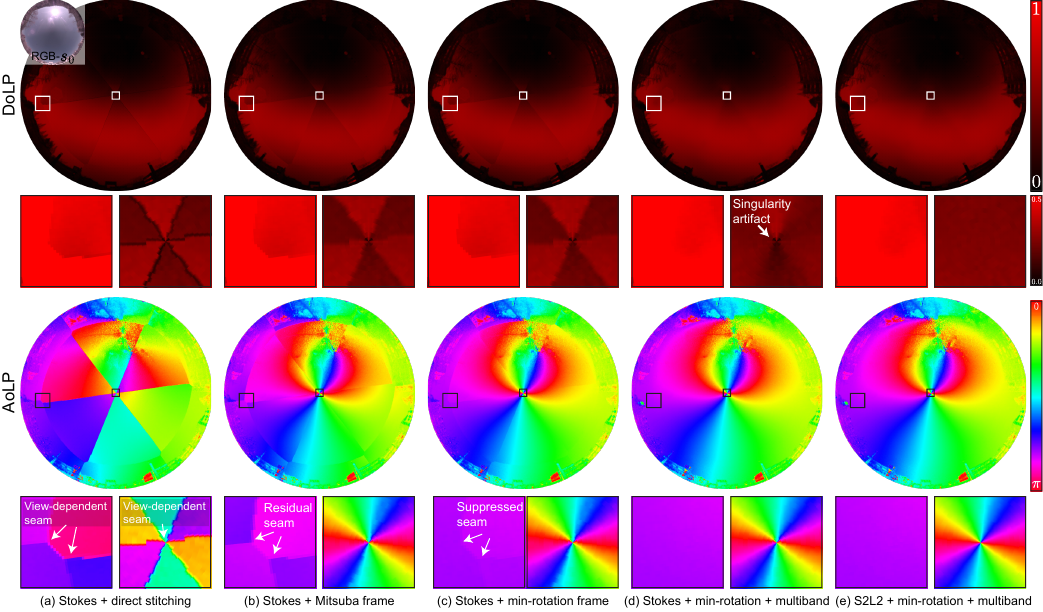}
    \caption{\textbf{Comparison of polarimetric panorama stitching strategies.}
    We compare stitched polarimetric environment maps obtained with different frame definitions and blending strategies. For visualization, the environment maps are projected onto a hemisphere. Each column shows DoLP and AoLP, with zoomed-in regions highlighting stitching seams (left patch) and the north pole (center patch). 
    (a) Directly stitching Stokes-vector images without frame correction produces severe viewpoint-dependent AoLP discontinuities. 
    (b) Using the perspective frame constructed from an auxiliary camera up vector, as in Mitsuba~3~\shortcite{Mitsuba3}, produces some continuity but still yields inconsistencies near image corners. 
    (c) The proposed minimum-rotation perspective frame improves consistency around the image corner.  
    Quantitatively, the mean/median AoLP angular errors in overlapping regions are $33.7^\circ/31.8^\circ$, $15.0^\circ/8.2^\circ$, and $12.4^\circ/3.3^\circ$ for (a), (b), and (c), respectively.
    (d) Applying multiband blending~\shortcite{burt1983multiresolution} directly to Stokes-vector images effectively smooths residual differences between views, but artifacts arise near frame singularities.
    (e) Blending in the S2L2 representation~\shortcite{yi2025frame} yields smooth and continuous DoLP and AoLP, avoiding both stitching seams and singularity-induced artifacts.}
    \label{fig:stitch_comparison}
\end{figure*}

We build our polarimetric panorama stitching pipeline upon the classical panorama formulation~\cite{szeliski1997creating,brown2007automatic}. In particular, established techniques such as image warping on the full-sphere domain and multiband blending~\cite{burt1983multiresolution} remain useful for spatial alignment and seamless composition. 
However, polarization imaging introduces an additional challenge: the Stokes vector is inherently frame-dependent, and therefore, unlike the intensity $s_0$ and the circular polarization $s_3$ components, the linear polarization components $(s_1, s_2)$ are frame-dependent: their values vary with the orientation of the transverse reference axes. 
As shown in Figure~\ref{fig:stitch_comparison}(a), directly applying a standard panorama stitching pipeline to Stokes-vector images produces visible discontinuities in the stitched panorama. 

To address this issue, we develop a polarimetric panorama stitching method that explicitly accounts for the frame-dependent nature of Stokes representations. Our method has two main components: (1) a perspective measurement frame that is suitable for wide-FoV polarimetric images, and (2) a singularity-aware blending strategy based on a continuous frame-free representation.


\begin{figure}[t]
    \centering
    \includegraphics[width=\hsize]{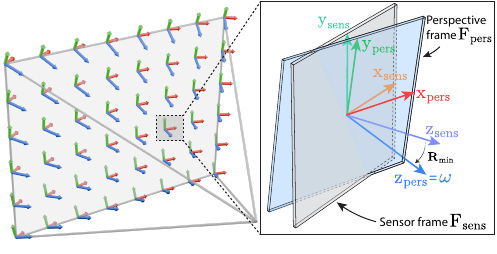}
    \caption{\textbf{Visualization of the perspective frame defined by the minimum-rotation construction.}
    Left: Perspective-frame field over the camera image plane. 
    Right: Magnified view of the frame at a single pixel.}
    \label{fig:minrot_pers_frame}
\end{figure}

\subsection{Measurement Frame under Perspective Projection}
Defining an appropriate reference frame for the Stokes vector is crucial, especially for wide-FoV cameras.
Under orthographic projection, the incident light direction is perpendicular to the sensor plane, and thus the transverse polarization basis is solely defined with the in-plane axes of the on-sensor micro-polarizers.
Under perspective projection, however, each pixel corresponds to a different propagation direction, and off-axis rays are not perpendicular to the sensor plane. Therefore, the Stokes components at each pixel should be interpreted with respect to a per-ray local measurement. This distinction becomes increasingly important toward the image periphery, where the incidence angle is larger.

Mitsuba~3~\cite{Mitsuba3} constructs a transverse frame using an auxiliary camera up vector. While this construction defines a valid frame perpendicular to the local ray direction, the fixed up-vector construction introduces a ray-dependent basis roll. For instance, two rays with the same polar angle but different azimuths can induce different roll angles. As shown in Figure~\ref{fig:stitch_comparison}(b), this frame reduces some discontinuities but still produces AoLP inconsistencies near image boundaries, where off-axis effects are more pronounced.

Instead of relying on a fixed auxiliary axis, we define the perspective frame by applying the minimum rotation from the sensor normal to each ray direction. Figure~\ref{fig:minrot_pers_frame} visualizes the notations and resulting frame. Let the local sensor frame be $\mathbf{F}_\mathrm{sens} = [\mathbf{x}_\mathrm{sens}, \mathbf{y}_\mathrm{sens}, \mathbf{z}_\mathrm{sens}]$, where $(\mathbf{x}_\mathrm{sens}, \mathbf{y}_\mathrm{sens})$ spans the sensor plane and $\mathbf{z}_\mathrm{sens}$ denotes the sensor-plane normal. Given the ray propagation direction $\boldsymbol{\omega}$ derived from the camera intrinsics at each pixel, we define the unique smallest-angle rotation $\mathbf{R}_\mathrm{min}\in \mathrm{SO}(3)$ that maps $\mathbf{z}_\mathrm{sens}$ to $\boldsymbol{\omega}$, i.e., $\mathbf{R}_\mathrm{min}\mathbf{z}_\mathrm{sens} = \boldsymbol{\omega}$. Specifically, we use the closed-form Rodrigues rotation
\begin{equation}
    \mathbf{R}_{\min}(\boldsymbol{\omega})
    = \mathbf{I}_3 + [\mathbf{v}]_\times +
    \frac{[\mathbf{v}]_\times^2}
         {1+\mathbf{z}_\mathrm{sens}^\intercal \boldsymbol{\omega}},
    \qquad
    \mathbf{v} = \mathbf{z}_\mathrm{sens} \times \boldsymbol{\omega},
\end{equation}
where $\mathbf{I}_3$ is the identity matrix and $[\mathbf{v}]_\times$ denotes the skew-symmetric matrix associated with $\mathbf{v}$.
While this equation is known as the standard shortest-arc rotation~\cite{moller1999efficiently}, we use it to construct a perspective frame field for Stokes-vector alignment. Accordingly, we 
define the minimum-rotation perspective frame $\mathbf{F}_\mathrm{pers}(\boldsymbol{\omega})= [\mathbf{x}_\mathrm{pers}, \mathbf{y}_\mathrm{pers}, \mathbf{z}_\mathrm{pers}]$ as a function of the ray propagation direction $\boldsymbol{\omega}$:
\begin{align}
    \mathbf{F}_\mathrm{pers}(\boldsymbol{\omega})
    &=
    \mathbf{R}_{\min}(\boldsymbol{\omega}) \mathbf{F}_\mathrm{sens} \\
    &= 
    \left[
    \mathbf{R}_{\min}(\boldsymbol{\omega}) \mathbf{x}_\mathrm{sens}, \
    \mathbf{R}_{\min}(\boldsymbol{\omega}) \mathbf{y}_\mathrm{sens}, \
    \boldsymbol{\omega}
    \right].
    \label{eq:frame_meas}
\end{align}
As shown in Figure~\ref{fig:stitch_comparison}(c), using the minimum-rotation perspective frame improves view consistency, particularly near image boundaries. This result indicates that the proposed frame definition better matches the actual measurement geometry and is suitable for wide-FoV polarimetric image stitching.







\begin{figure*}[t]
  \includegraphics[width=\linewidth]{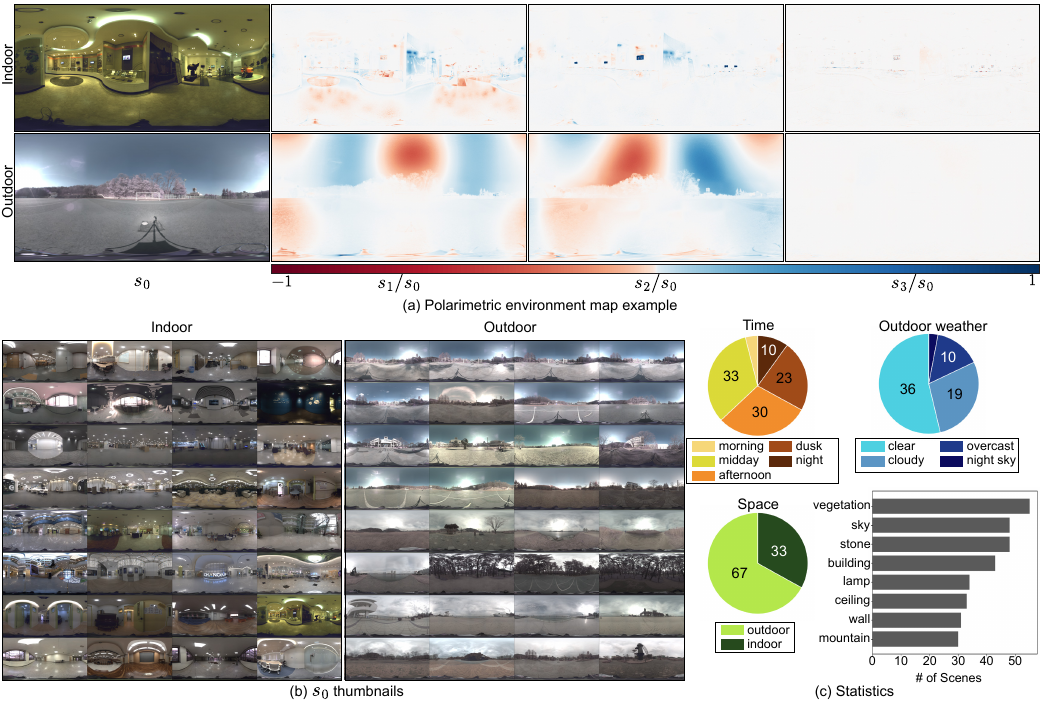}
  \caption{\textbf{Overview of our polarimetric environment map dataset.} (a) Representative polarimetric environment maps captured in indoor and outdoor scenes. (b) Thumbnail images of indoor and outdoor scenes, illustrating the environmental diversity of our dataset. (c) Dataset statistics summarizing the distribution of capture times, outdoor weather conditions, scene types, and semantic categories.}
  \label{fig:dataset}
\end{figure*}

\subsection{Image blending on the S2L2.}

Even with an appropriate perspective measurement frame, singularities are unavoidable when representing tangent-frame quantities over the sphere~\cite{yi2024spin,yi2025frame}. Near such singularities, the physical polarization field may be smooth, but its Stokes components can exhibit abrupt coordinate-induced variations. Directly blending these Stokes components can therefore introduce visible artifacts. As shown in Figure~\ref{fig:stitch_comparison}(d), these artifacts become particularly pronounced when multiband blending~\cite{burt1983multiresolution} is applied directly to Stokes-vector images.

To avoid this issue, we perform blending in the S2L2 representation~\cite{yi2025frame} rather than directly in the Stokes-vector representation. S2L2 represents the linear polarization components $(s_1, s_2)$ using ten continuous components over the sphere, avoiding explicit dependence on a singular tangent-frame parameterization. To our knowledge, we are the first to exploit this property for seamless polarimetric image blending.

In our pipeline, we first convert each Stokes-vector image into the S2L2 representation. The image contains three RGB channels and twelve polarization channels: two channels for $s_0$ and $s_3$, and ten channels for the S2L2 representation of $(s_1, s_2)$. We then warp all channels into the equirectangular panorama domain and blend the warped views using multiband blending. As shown in Figure~\ref{fig:stitch_comparison}(e), blending in the S2L2 representation produces smooth and continuous polarization states while avoiding both stitching seams and singularity-induced artifacts.

\section{Polarimetric Environment-map Dataset}
\label{sec:penvmap}
\paragraph{Data Format}
Using the proposed imaging system and polarimetric panorama stitching method, we capture polarimetric environment maps for 100 scenes. The dataset also includes a time-varying sequence of 10 polarimetric environment maps acquired over the course of a day from a fixed outdoor location. Each environment map is stored as an equirectangular panorama image with a resolution of 2048$\times$1024, storing per-pixel Stokes vector for each RGB channel. Figure~\ref{fig:dataset}(a) shows representative thumbnails of the intensity component ($s_0$) and normalized polarization components ($s_1/s_0$, $s_2/s_0$, $s_3/s_0$).
\paragraph{Scene Diversity}
As summarized in Figure~\ref{fig:dataset}(b), the dataset includes a wide range of real-world environments, including 33 indoor and 67 outdoor scenes. Indoor scenes comprise offices, laboratories, corridors, and public interior spaces, which contain diverse artificial lighting configurations and complex inter-reflection effects. Outdoor scenes include both natural and urban environments, such as parks, roads, plazas, and beaches.

\paragraph{Time and Weather}
We also provide auxiliary metadata for each scene, including time of day and weather conditions. Time of day is categorized into morning, midday, afternoon, dusk, and night, thereby capturing variations in solar elevation and low-light conditions. Weather conditions are labeled as clear or cloudy for daytime scenes, and as night sky for nighttime captures.

\section{Polarimetric Rendering}
\label{sec:polarimetric_rendering}

To enable rendering with our dataset, we use the physics-based renderer Mitsuba~3~\cite{Mitsuba3}, which supports simulation of polarized light transport. We implement a custom renderer plugin by extending the default environment emitter to support our polarized environment map.

Figure~\ref{fig:teaser} shows polarimetric rendering results obtained using our polarimetric environment maps. In contrast to conventional radiance-only environment maps, our polarimetric environment maps produce rendered images exhibiting spatially varying polarization effects that arise naturally from environmental lighting. These results demonstrate that incorporating polarization into environment illumination is essential for physically grounded image synthesis and for modeling complex polarization patterns in real-world lighting.

\section{Assessment}
\label{sec:assessment}

\subsection{Comparison between Real and Synthetic Images}

We evaluate the physical fidelity of our polarimetric environment maps by comparing real Stokes-vector images with synthetic images rendered in Mitsuba~3.
For the real-world reference, we capture a metallic sphere using the Stokes camera under a clear sky. We also acquire a polarimetric environment map at the same location and time.
To reproduce this configuration synthetically, we construct a corresponding scene in Mitsuba~3. 
Since exactly matching the geometry and pBRDF of the real object is impractical, we use an ideal sphere with a rough conductor material and grid-search plausible material parameters using the $s_0$ appearance. We also configure the synthetic camera using the calibrated intrinsic and extrinsic parameters of the real camera.

\begin{figure*}[t]
    \includegraphics[width=\linewidth]{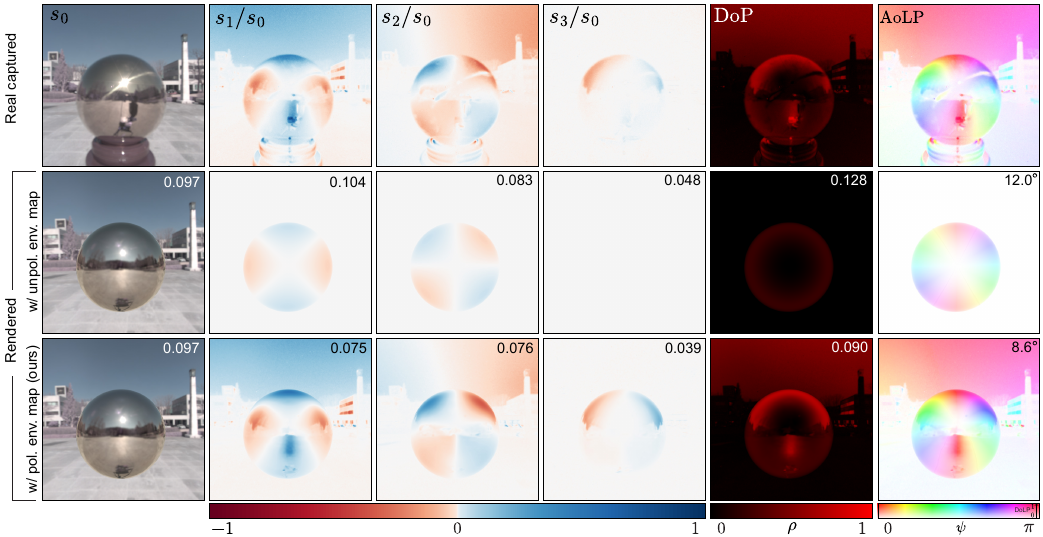}
    \caption{
        \textbf{Qualitative comparison between real captures and rendered results.}
        Top: We capture a real metallic sphere under a clear sky and simultaneously acquire a polarimetric environment map at the same location and time.
        Middle: We render a synthetic image of a metallic sphere using the captured environment map while assuming unpolarized illumination, which uses only $s_0$, and $s_1$, $s_2$, and $s_3$ are set to zero. Because this assumption neglects the real-world polarization state of incoming light, the resulting polarization deviates from real observation.
        Bottom: In contrast, rendering with our polarimetric environment map yields results that closely match the real captured appearance. 
        Values in the upper-right corner of each rendered panel report the mean absolute error relative to the real capture, with AoLP errors given in degrees.}
        
    \label{fig:assessment_sphere1}
\end{figure*}

Figure~\ref{fig:assessment_sphere1} shows qualitative comparisons between the real captures and the rendered results. The first row presents the real captured images, whereas the second and third rows show the corresponding renderings.
To assess the effect of polarized illumination, we construct an unpolarized variant of the captured environment map by retaining only $s_0$ and setting $s_1$, $s_2$, and $s_3$ to zero. Rendering with this unpolarized environment map fails to reproduce the polarization characteristics observed in the real capture.
In contrast, rendering with our polarimetric environment map produces Stokes vector images that closely match the real observations across all Stokes parameters. Specifically, regions reflecting buildings exhibit relatively weak polarization responses, whereas regions of sky and sunlit ground produce stronger responses. In addition, the real capture contains a circular polarization component $s_3$. This component is absent when using the unpolarized environment map but is faithfully reproduced when rendering with our polarimetric environment map. Additional results under different environment maps are provided in the supplementary material.


\subsection{Validation under Wide Field of View}
Our imaging system employs a wide-FoV lens with approximately $100^\circ$ FoV. 
Toward the image corners, such optics can introduce undesirable geometric and polarimetric artifacts, including radial distortion, vignetting, and non-perpendicular incidence on the image sensor. Our pipeline accounts for these effects through dedicated calibration and image stitching.

\begin{figure*}[t]
    \centering
    \includegraphics[width=\linewidth]{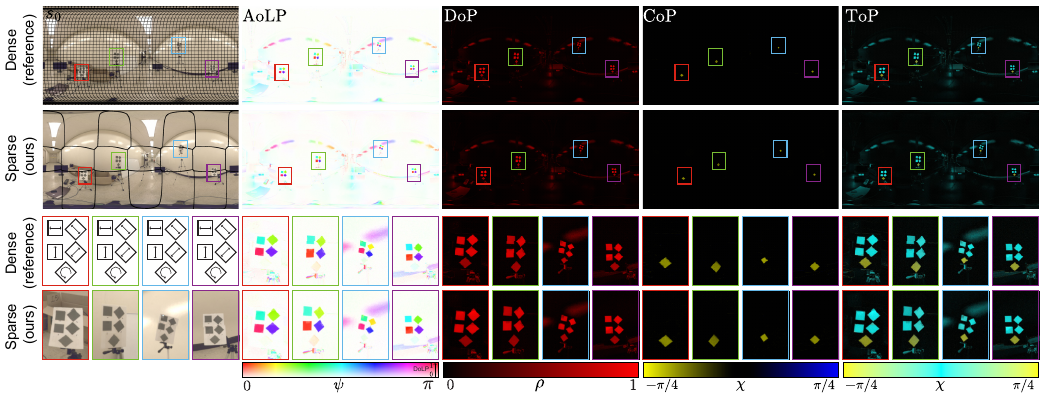}
    \caption{\textbf{Comparison between the densely scanned reference and our sparsely acquired reconstruction.}
    We compare a densely scanned reference acquired from 1860 images (Dense) with our sparse acquisition using only 18 images (Sparse). The black lines overlaid on the $s_0$ images indicate the boundaries of the dominant contributing view region. The scene includes four patches consisting of five polarization targets: four linear polarizers oriented at approximately $0^\circ$, $45^\circ$, $90^\circ$, and $135^\circ$, and one left-handed circular polarizer. 
    Since the target polarization axes were not geometrically calibrated, absolute polarimetric evaluation is infeasible; we therefore assess the sparse reconstruction against the dense reference. The resulting MAEs for $s_1/s_0$, $s_2/s_0$, and $s_3/s_0$ are $0.0226$, $0.0284$, and $0.0113$, respectively, demonstrating close agreement between the sparse and dense reconstructions.}
    
    \label{fig:scanning}
\end{figure*}

To validate the effectiveness of these components, we compare the polarimetric environment map reconstructed by our method against a reference map obtained from densely sampled viewpoints. The reference map is acquired by rotating the camera at $6^\circ$ angular intervals, resulting in 1860 captures in total. For each view, we use only the central 125$\times$125-pixel crop, where optical degradation is minimal, to construct a high-quality reference.
Figure~\ref{fig:scanning} compares this dense-scan reference environment map with the map reconstructed using our sparse 18-view acquisition.
Despite the substantially sparser acquisition, our reconstruction closely agrees with the dense-scan reference, demonstrating the consistency of our sparse reconstruction with the dense-scan reference under wide-FoV capture.


\subsection{Sky Polarization}

\begin{figure*}[t]
    \centering
    \includegraphics[width=\linewidth]{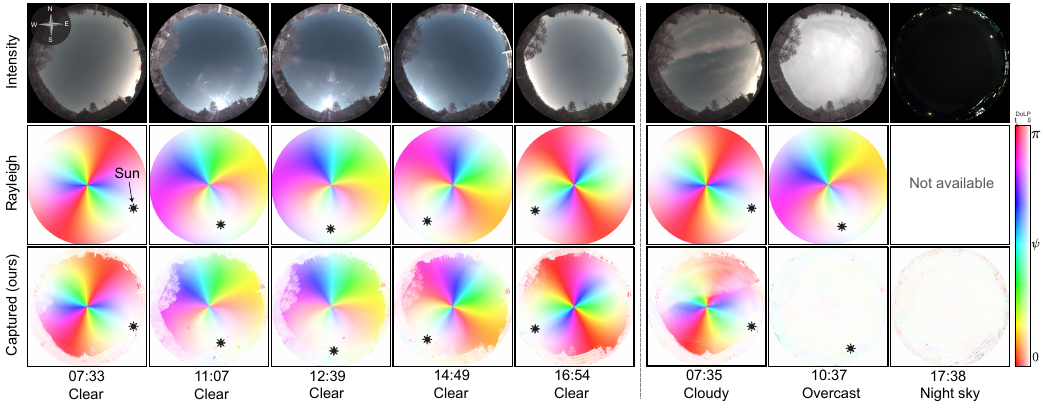}
    \caption{
    \textbf{Real-world polarization states of the sky.}
    We compare the AoLP and DoLP in our captured data with those computed by the Rayleigh scattering model~\shortcite{strutt1871lviii, strutt1871xv}. Left: Under clear-sky conditions, the polarization pattern varies over time according to the sun's position and closely follows the Rayleigh scattering model. Right: By contrast, under non-clear-sky conditions, our dataset includes realistic polarization states that cannot be represented by simple analytic models, including effects induced by clouds and the night sky.}
    \label{fig:sky}
\end{figure*}

\paragraph{Time Dependency.}
Under clear-sky conditions, polarization patterns are strongly governed by solar motion~\cite{miyazaki2009estimating,wilkie2004analytical}. To analyze this behavior, our dataset includes a time-varying sequence of polarimetric environment maps captured from fixed viewpoints, comprising 10 maps acquired throughout the day.
Figure~\ref{fig:sky}(left) shows representative clear-sky sequences from early morning to late afternoon. We compare these measurements with predictions from the Rayleigh scattering model~\cite{strutt1871lviii,strutt1871xv}, computed from the corresponding sun position. The measured polarization states closely follow the model predictions: the AoLP exhibits a coherent, sky-wide pattern that rotates smoothly as the sun moves from east to west. This result confirms that our measurements are consistent with the classical scattering model under clear-sky conditions.

\paragraph{Weather Dependency.}
Sky polarization is highly sensitive to atmospheric conditions, including clouds, overcast skies, and nighttime illumination. Figure~\ref{fig:sky}(right) shows polarimetric environment maps captured under diverse weather conditions. In contrast to clear-sky observations, cloudy and overcast skies exhibit reduced DoLP around dense cloud regions, owing to the depolarizing effect of multiple scattering in clouds~\shortcite{hegedus2007polarization, pust2008digital}. At night, the absence of direct solar illumination produces low DoLP across the entire sky. Together, these measurements demonstrate that real-world sky polarization varies with atmospheric conditions, motivating weather-aware modeling and analysis.

\subsection{Shape from Polarization}

Our polarimetric environment maps enable controlled evaluation of polarimetric imaging methods under diverse radiance and polarization conditions. We demonstrate this capability using Shape from Polarization (SfP) as an example. Classical model-based SfP methods often assume unpolarized illumination~\cite{miyazaki2003polarization, smith2018height}, while Shape from Sky~\cite{ichikawa2021shape} explicitly models clear-sky polarization under Rayleigh scattering~\shortcite{strutt1871lviii,strutt1871xv}; both are therefore tied to specific illumination assumptions. Learning-based methods~\cite{ba2020deep, lyu2024sfpuel} improve robustness to diverse radiance distributions but typically retain the unpolarized-illumination assumption. This motivates us to use SfP as a case study for a controlled comparison between radiance-only and polarimetric illumination.

\begin{figure*}[t]
    \centering
    \includegraphics[width=\linewidth]{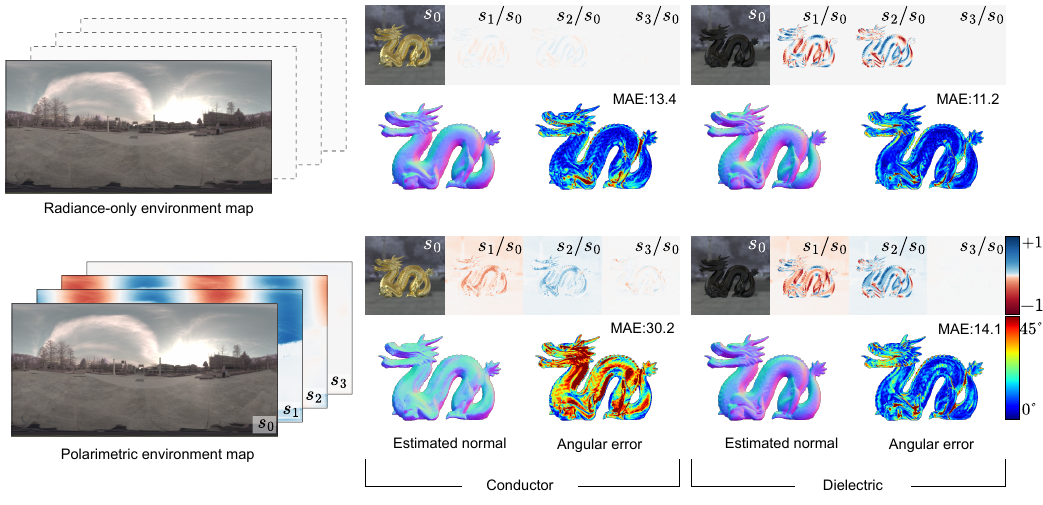}
    \caption{
        \textbf{Impact of polarimetric environment maps on SfP.}
        We evaluate a pretrained SfPUEL model~\shortcite{lyu2024sfpuel} under two illumination settings: a conventional radiance-only environment map, which assumes unpolarized illumination, and our polarimetric environment map.
        Although the input radiance is identical, the sky polarization changes the rendered Stokes observations and violates the unpolarized-illumination assumption used by SfPUEL.
        This comparison shows the potential impact of real-world polarized illumination on SfP and demonstrates how our dataset enables controlled experiments that isolate polarization effects.
    }
    \label{fig:shape_from_polarization}
\end{figure*}

Figure~\ref{fig:shape_from_polarization} compares pretrained SfPUEL~\shortcite{lyu2024sfpuel} under radiance-only and polarimetric renderings of otherwise identical scenes. The predicted normals differ between the two illumination settings, with a larger change for the conductor, consistent with its stronger preservation of incident polarization. While this limited comparison does not support a general conclusion about SfPUEL's robustness to polarized illumination, it demonstrates how our dataset enables controlled experiments that isolate polarization effects from radiance.

\section{Conclusion}
\label{sec:conclusion}

We present a real-world dataset of polarimetric environment maps, comprising 100 scenes captured under diverse indoor and outdoor illumination conditions. To construct this dataset, we develop a dedicated acquisition system that combines a Stokes camera with a motorized rotation stage, together with a polarization-aware panorama-stitching method. Our stitching pipeline produces smooth, continuous, and artifact-free polarimetric panoramas by combining minimum-rotation measurement-frame construction with an S2L2 representation for singularity-aware blending. 
By enabling physically faithful polarization-aware light transport simulation, systematic analysis of polarimetric illumination, and realistic synthetic polarimetric data generation, our dataset provides a foundation for future polarization-based methods in inverse rendering, material and shape estimation, and broader polarimetric visual computing applications.

\paragraph{Discussion.} 
Although our dataset provides full-Stokes environment maps, we observe that the circular polarization component $s_3$ is negligible in most scenes. This suggests that natural environmental illumination is predominantly unpolarized or linearly polarized, while circular polarization occurs only rarely. Consequently, in many practical scenarios under natural illumination, a linear polarization camera may be sufficient for acquiring polarimetric environment maps.

\paragraph{Limitation.}
One characteristic of our current acquisition setup is that, although the Stokes camera used in this work is primarily designed for the visible spectrum, the sensor exhibits residual sensitivity to near-infrared (NIR) light. As a result, in some scenes, particularly outdoor scenes, NIR radiation from sunlight is partially recorded in the red channel, which can affect perceptual color reproduction and the red-channel Stokes vector.





\begin{acks}
This work was partly supported by the National Research Foundation (NRF) grants
(RS-2023-00211658 and RS-2024-00438532), and the Institute of Information \&
Communications Technology Planning \& Evaluation (IITP) grants
(RS-2024-0045788; IITP-2026-RS-2024-00437866, Information Technology Research
Center; and RS-2026-25616370, AI Star Fellowship Support Program), funded by
the Korean government (MSIT).
\end{acks}

\bibliographystyle{ACM-Reference-Format}
\bibliography{./references}



  

\end{document}